\documentclass[conference]{IEEEtran}

\usepackage{url}
\usepackage{graphicx}
\usepackage{subfig}
\usepackage{enumitem}
\usepackage{verbatim}
\usepackage{multicol}

\begin{document}

\title{RBioCloud: A Light-weight Framework for\\Bioconductor and R-based Jobs on the Cloud}

\author	{
	\IEEEauthorblockN{Ishan Patel}
	\IEEEauthorblockA{IBM Canada Ltd.\\
		50 Innovations Drive\\
		Bedford, NS, Canada\\
		Email: ishanp@ca.ibm.com}
	\and
	\IEEEauthorblockN{Blesson Varghese\textsuperscript{1} and Adam Barker}
	\IEEEauthorblockA{Big Data Laboratory (\url{http://bigdata.cs.st-andrews.ac.uk})\\
		School of Computer Science, University of St Andrews\\
		St Andrews, Fife, UK\\
		Email: \{varghese, adam.barker\}@st-andrews.ac.uk}
	}

\maketitle

\footnotetext[1]{Corresponding Author (\url{http://www.blessonv.com})}

\begin{abstract}
Large-scale ad hoc analytics of genomic data is popular using the R-programming language supported by 671 software packages provided by Bioconductor. More recently, analytical jobs are benefitting from on-demand computing and storage, their scalability and their low maintenance cost, all of which are offered by the cloud. While Biologists and Bioinformaticists can take an analytical job and execute it on their personal workstations, it remains challenging to seamlessly execute the job on the cloud infrastructure without extensive knowledge of the cloud dashboard. How analytical jobs can not only with minimum effort be executed on the cloud, but also how both the resources and data required by the job can be managed is explored in this paper. An open-source light-weight framework for executing R-scripts using Bioconductor packages, referred to as `RBioCloud', is designed and developed. RBioCloud offers a set of simple command-line tools for managing the cloud resources, the data and the execution of the job. Three biological test cases validate the feasibility of RBioCloud. The framework is publicly available from \texttt{\url{http://www.rbiocloud.com}}.
\end{abstract}


\IEEEpeerreviewmaketitle

\section{Introduction}
Ad-hoc analytics of genomic data is popular in domains such as computational biology and bioinformatics. Typically, an analytical job comprises software scripts written by biologists or bioinformaticists in high-level programming languages, such as R \cite{Rprogramming}, along with large amounts of data that needs to be processed. R-based analytics in computational biology or bioinformatics is gaining popularity and is supported through 671 software packages provided by Bioconductor \cite{bioconductor}. Analytical jobs which may require a few hours or perhaps even a few days may ingest large amounts of data and subsequently also produce data in large volumes. Not only is analytics inherently computationally intensive, but also data intensive. High-performance computing systems have therefore become attractive for executing large-scale analytical jobs \cite{prelim10}. 

Traditional high-performance computing systems such as clusters and supercomputers offer a good platform to perform large-scale analytics. However, it is required of the computational biologist and bioinformaticist, who has excellent programming and statistical skills, to also have extensive knowledge of the high-performance computing hardware. Moreover, the costs required for investing in large-scale systems and their maintenance is high. The cloud has become an appealing high-performance computing platform for ad-hoc analytics since it offers on-demand computing and storage resources, along with scalability and low maintenance costs \cite{prelim2, prelim7, prelim5-2}. This has led to a variety of research for supporting computational biology and bioinformatics related jobs on the cloud (for example, genome sequencing \cite{prelim2-1}, genome informatics \cite{prelim8}, comparative genomics \cite{prelim3-1}, proteomics \cite{prelim4-1} and biomedical computing \cite{prelim5-1}).

Software projects such as elasticR \cite{elasticR}, DARE \cite{DARE} and AzureBlast \cite{AzureBlast} support applications on the cloud, all of which require the user to have extensive knowledge of the cloud dashboard to be able to port an existing analytical workload onto the cloud. The options provided by such projects for a fully configurable cloud cluster can fit well with the skill set of a cloud developer, thereby narrowing their wide usage. The major challenge in the research of developing software similar to the ones above for Computational Biology and Bioinformatics (for example, CloudBLAST \cite{CloudBLAST}, GalaxyCloudMan \cite{GalaxyCloudMan}, SIMPLEX \cite{SIMPLEX} and Crossbow \cite{Crossbow}) is to seamlessly execute an analytical job on the cloud in a manner similar to how the job would be executed on the personal workstation. However, the use of such software adds an additional layer of complexity for managing the software on top of executing the job. Further, adapting the above projects to execute workloads developed using the R programming language is cumbersome, specific adaptations being required in many cases. A similar challenge exists for executing the increasing number of analytical workloads that are developed using the R with Bioconductor packages \cite{RandBioconductor}. 

The current Bioconductor based solution \cite{BioconductorCloudAMI} is based on manually configuring the cloud dashboard for every job that needs to be executed. Software developed to support R and Bioconductor, for example, Myrna \cite{Myrna}, Contrail \cite{Contrail} and \cite{Jnomics} are restricted to specific applications in computational biology and bioinformatics. These challenges can be overcome by the development of a generic framework that can support R-based jobs supported by Bioconductor packages, and their execution and management on the cloud.

The research reported in this paper aims to address the above challenges. A light-weight framework, `RBioCloud', for supporting R-based analytical applications which use Bioconductor software packages and need to be executed on the cloud is presented. Domain scientists have to often spend a lot of time dealing with the complex details of configuring the cloud. Using RBioCloud, an analytical job can be executed on the cloud with minimal effort using a set of five commands from a personal workstation. The need for any extensive knowledge of the cloud dashboard is minimised.

The contributions of RBioCloud research is a framework (i) for handling a diverse range of Bioconductor based analytical jobs on the cloud, (ii) for abstracting the complexities of cloud set up and configuration, (iii) for computational biologist and bioinformaticists to easily access and use the cloud, thereby saving time, and (iv) with seemingly minimal difference between a domain scientists workstation though remote resources are accessed. The feasibility of RBioCloud is validated by using three test cases employing Bioconductor packages for executing R-based scripts on the cloud. In the first test case, genome searching is performed on a single cloud instance, in the second test case, differentially expressed genes are detected on a single cloud instance, and in the third test case, normalisation of microRNA (miRNA) microarray data is performed on a cluster in the cloud.

The remainder of this paper is organised as follows. Section \ref{design} considers the design of the RBioCloud framework. Section \ref{tools} describes the command line tools offered by RBioCloud for managing and executing an analytical job. Section \ref{feasibilitystudy} presents three test cases to validate the feasibility of RBioCloud. Section \ref{conclusions} concludes this paper by considering future work. 

\section{Framework Design}
\label{design}
\begin{figure}
	\centering
		\includegraphics[width=0.33\textwidth]{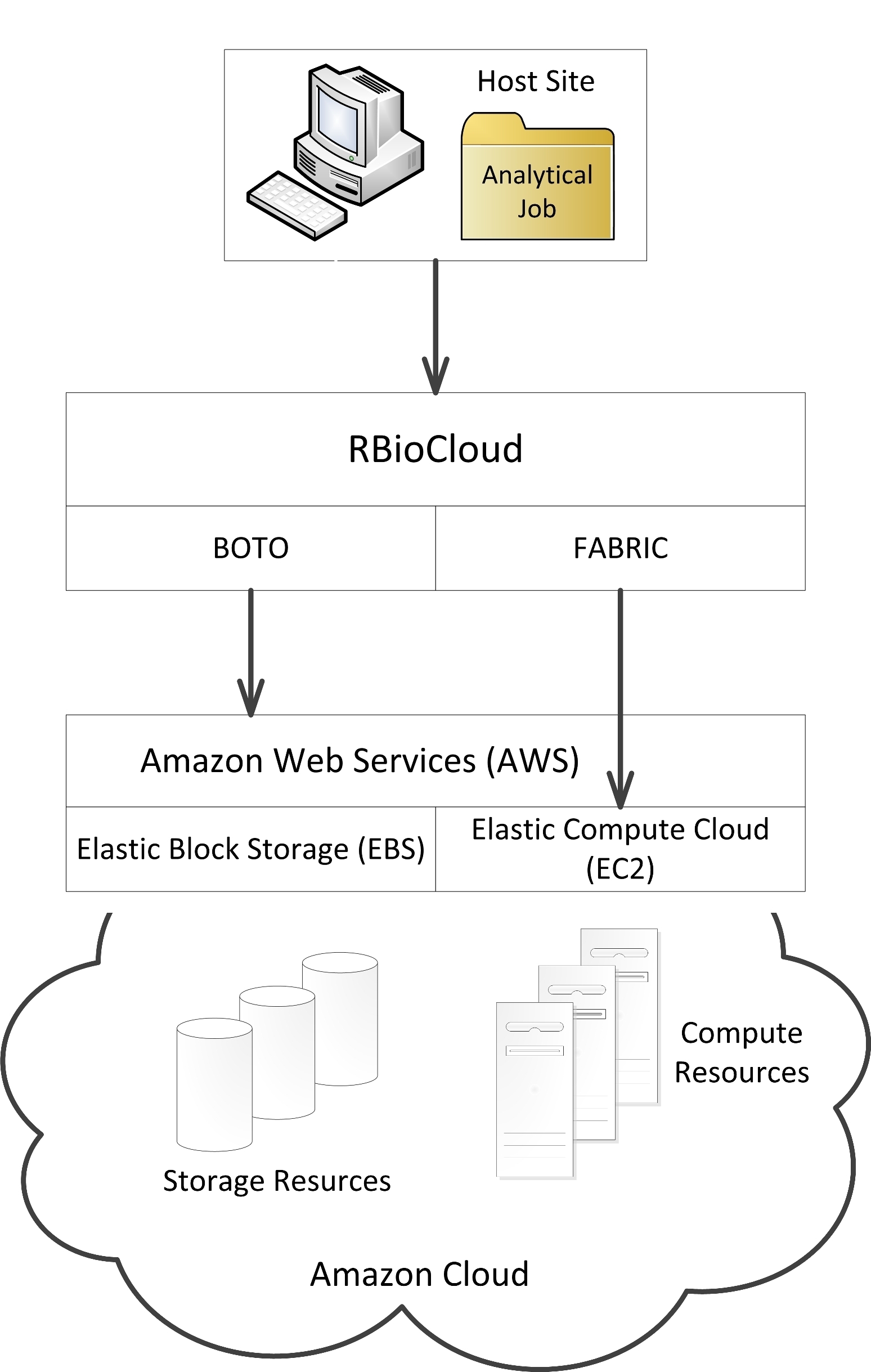}
	\caption{Design of RBioCloud framework}
\label{figure1}
\end{figure}

\begin{figure*}
	\centering
		\includegraphics[width=0.93\textwidth]{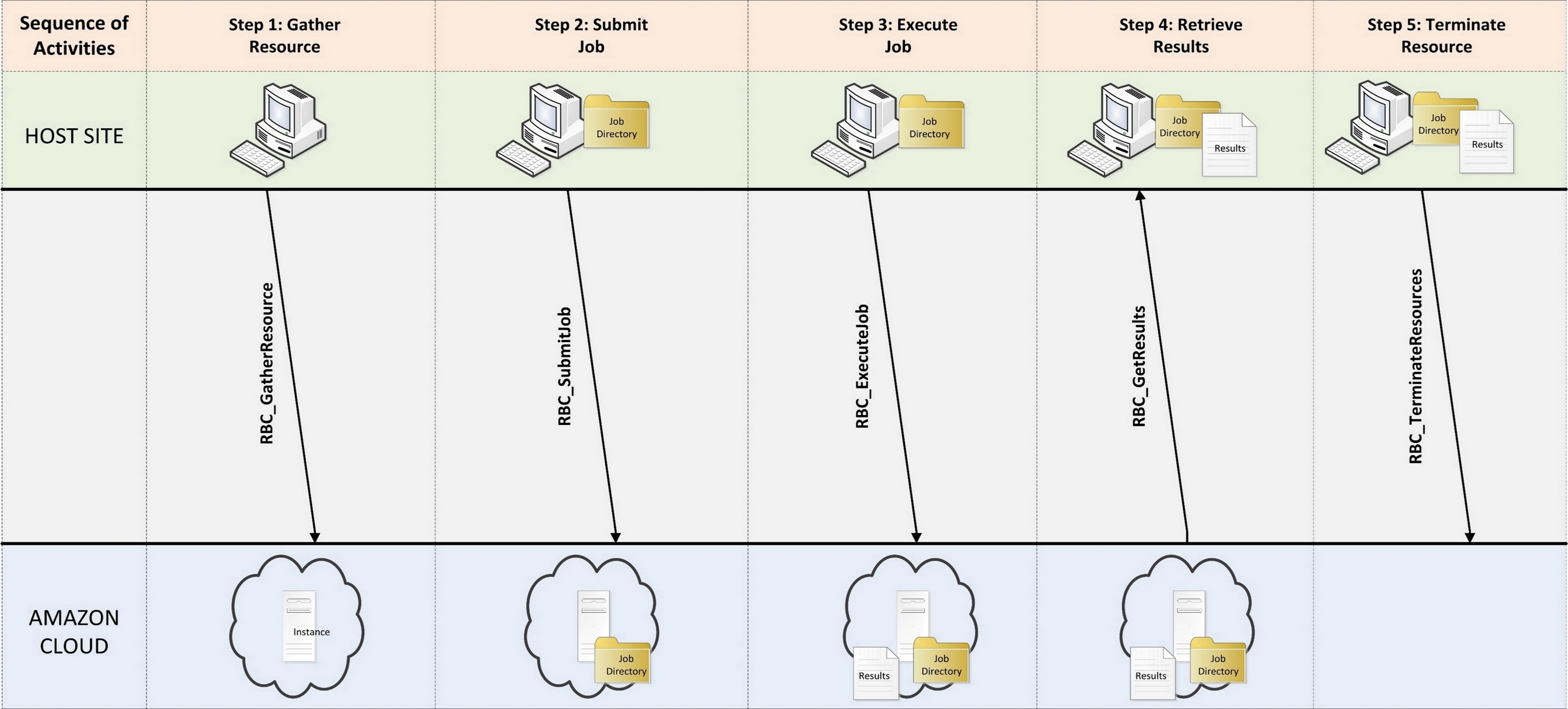}
	\caption{Sequence of activities while using the RBioCloud framework}
\label{figure2}
\end{figure*}

Figure \ref{figure1}, shows the design of the RBioCloud framework which is located on a host site for accessing and managing cloud resources. The host site represents the workstation of a computational biologist or a bioinformaticist who makes use of the cloud infrastructure to execute a job. The Amazon cloud infrastructure is employed in this research. RBioCloud is designed so that the job can be executed from the host site using the following five step sequence (refer Figure \ref{figure2}): 

\begin{itemize}
	\item \textit{Step 1:} Gather resources - initialise cloud compute and storage resources from the host.
	\item \textit{Step 2:} Submit job - send the analytical job from the host onto cloud resources.
	\item \textit{Step 3:} Execute job - execute the scripts within the job on the resources.
	\item \textit{Step 4:} Retrieve results - get results generated on the cloud resources onto host.
	\item \textit{Step 5:} Terminate resources - release all resources which were initialised on the cloud. 
\end{itemize}

\subsection{Supporting Interfaces}
RBioCloud is developed using the Python programming language and is supported by a number of interfaces. The compute and storage resources are provided by the Amazon Web Services (AWS)\footnote{\small \url{http://aws.amazon.com/}}. All resources are available on-demand and are paid for on the basis of their usage. The computational resources are offered through Elastic Compute Cloud (EC2)\footnote{\small \url{http://aws.amazon.com/ec2/}} and are available as instances. The storage resources are referred to as the Elastic Block Storage (EBS)\footnote{\small \url{http://aws.amazon.com/ebs/}} provide persistent data storage. Two Python interfaces, namely BOTO\footnote{\small \url{https://github.com/boto/boto}} provides the interface to access the resources provided by AWS and Fabric\footnote{\small \url{http://docs.fabfile.org/en/1.4.3/}} facilitates remote administration of the cloud resources.

Amazon instances are initialized using Amazon Machine Images (AMI)\footnote{\small \url{http://aws.amazon.com/amis}}. The RBioCloud framework is built on the Bioconductor Cloud AMI \cite{BioconductorCloudAMI} and supports the R programming language along with Bioconductor packages. 

The cloud is attractive for large analytical jobs as parallel computations incorporated within jobs can be exploited on the cloud. The Simple Network Of Workstations (SNOW)\footnote{\small \url{http://www.sfu.ca/\textasciitilde sblay/R/snow.html}} interface is employed for parallel execution of jobs on the cloud. 

\section{Tools}
\label{tools}
The five Command line tools offered by RBioCloud to support gathering of cloud resources, to submit and execute a job, retrieve results from the cloud and terminate resources are presented in this section. 

\subsection{Gather Resources}
\texttt{RBC\_GatherResource} provisions configuring an instance or multiple instances and a cluster on the cloud. The syntax of the command is 
\begin{enumerate}[leftmargin=10pt]
\item[]\texttt{RBC\_GatherResource [-h] [-v] [-rname RESOURCE\_NAME] [-rsize RESOURCE\_SIZE] [-ebsvol EBS\_VOLUME | -snap EBS\_SNAP] [-type INSTANCE\_TYPE] [-desc RESOURCE\_DESCRIPTION]}
\end{enumerate}

The optional arguments are: (a) \texttt{rname} to name a resource (instance or cluster) that is created, (b) \texttt{rsize}, to specify the size of the resource (if size is one, then only one instance is created, else if size is greater than one, then the instances are configured as a cluster), (c) \texttt{ebsvol} and \texttt{snap}, which are not specified at the same time. \texttt{ebsvol} specifies the EBS volume ID when an EBS volume is created. \texttt{snap} specifies the EBS snapshot ID from which an EBS volume can be created. \texttt{ebsvol} can be specified when an EBS volume is available, however, if \texttt{snap} is specified then a new EBS volume is created from the snapshot specified. If both arguments are not provided, then a default snapshot from a configuration file is used, (d) \texttt{type}, which defines the Amazon EC2 instance type\footnote{\small \url{http://aws.amazon.com/ec2/instance-types/}} based on the computational requirements of the task, and (e) \texttt{desc}, which can be used to provide a description for a resource.

\subsection{Submit Job}
A job comprises the script that needs to be executed and the data required by the script both of which need to be submitted to the cloud. The `rsync' protocol is used to submit the job. One advantage of using rsync is that subsequent data transfers are quickly synchronised between the host and the cloud. The submission of a job is facilitated using \texttt{RBC\_SubmitJob} and the syntax is 
\begin{enumerate}[leftmargin=10pt]
\item[]\texttt{RBC\_SubmitJob [-h] [-v] [-rname RESOURCE\_NAME [-toallnodes | -tomaster]] [-jobdir JOB\_DIRECTORY][-data]}
\end{enumerate}

The optional arguments are: (a) \texttt{rname} to specify the resource to which the job needs to be submitted. If a resource is not specified then the default resource from RBioCloud's configuration file is employed, (b) \texttt{jobdir} to specify the job directory at the host. If the job directory is not specified then the current working directory at the host site is considered to be the source job directory. The destination job directory is not provided since in the current setup the host job directory is synchronised to the home directory of the root user on the cloud. The job directory comprises a set of R scripts, a set of data files required by the scripts and a sub-directory that will contain results after the execution of the script. 

The optional switch \texttt{-tomaster} (default) submits the job to the master node of a cluster, while \texttt{-toallnodes} submits the job to all nodes of a cluster. The \texttt{-data} switch synchronises any folder not adhering to the structure of the job directory on to the resource. 

\subsection{Execute Job}
\texttt{RBC\_ExecuteJob} executes a job on the cloud resource. This command locks the resource onto the job and is only available for any additional use after the job has completed. The syntax of the command is
\begin{enumerate}[leftmargin=10pt]
\item[]\texttt{RBC\_ExecuteJob [-h] [-v] [-rname RESOURCE\_NAME] [-jobdir JOB\_DIRECTORY] [-rscript R\_SCRIPT] [-runname RUN\_NAME]} 
\end{enumerate}

The optional arguments of are: (a) \texttt{rname} to specify the resource on which the job needs to be executed, (b) \texttt{jobdir} indicates the job directory at the host site; the job with the same name from the corresponding job directory on the cloud is executed, and (d) \texttt{rscript} to indicate the R script to be executed. If \texttt{rscript} is not provided then the user is prompted to select from a list of R scripts that are available in the job directory.

The mandatory argument \texttt{runname} specifies the name of a run to distinguish multiple executions of a particular job.

\subsection{Retrieve Results}
\texttt{RBC\_GetResults} retrieves results from the cloud resource onto the host and the syntax is
\begin{enumerate}[leftmargin=10pt]
\item[]\texttt{RBC\_GetResults [-h] [-v] [-rname RESOURCE\_NAME [-frommaster | -fromall]] [-jobdir JOB\_DIRECTORY] [-runname RUN\_NAME]} 
\end{enumerate}

The optional arguments are: (a) \texttt{rname} to specify the resource from where the results need to be retrieved, (b) \texttt{jobdir} to indicate the location of the source job directory at the host site; the results are fetched from the corresponding job directory on the cloud. If no job directory is specified then the current working directory at the host site is used. 

The mandatory argument \texttt{runname} indicates the name of a run that was specified during execution and whose results need to be gathered. This argument can be used when the same R script has been executed a number of times and each execution had to be differentiated. Within the job directory the results are generated in a sub-directory. There are two scenarios of generating results on a cluster. In the first scenario, the master instance aggregates results from all worker instances and stores them on the master instance, and retrieval from the master instance is possible using \texttt{-frommaster}. In the second scenario, the results are generated on all instances, and retrieving results is possible using \texttt{-fromall}. 

\subsection{Terminate Resources}
After the completion of a job, the resources on the cloud need to be safely released to avoid billing of unused resources. \texttt{RBC\_TerminateResource} facilitates this and the syntax is
\begin{enumerate}[leftmargin=10pt]
\item[]\texttt{RBC\_TerminateResource [-h] [-v] [-rname RESOURCE\_NAME] [-deletevol]}
\end{enumerate}

The optional arguments are: (a) \texttt{rname} to specify the resource that needs to be terminated. The optional switch \texttt{-deletevol} deletes the EBS volume attached to the resource being terminated.

All the above commands can be used with two switches; firstly, \texttt{-h} to provide a description of the use and arguments of the command, and secondly, \texttt{-v} to provide provides the version of the installation.

\section{Feasibility Study}
\label{feasibilitystudy}
Popular biological jobs include searching, analysing and normalising data \cite{biotasks-1}. In this section three test cases that represent such biological jobs are selected to demonstrate the feasibility of RBioCloud for Bioconductor and R based jobs. Firstly, genome searching, secondly, detecting differential expression of genes and thirdly, normalisation of microRNA (miRNA) microarray data are presented. In the first and second test cases a single Amazon EC2 instance is used while in the third test case a cluster of Amazon EC2 instances are employed.

\subsection{Test case 1: Genome searching on an Instance}
The first test case is based on the \texttt{BSgenome} software package \cite{workflow1} available from Bioconductor\footnote{\small \url{http://www.bioconductor.org/packages/release/bioc/}}, and the script executed is \texttt{GenomeSearching.R} which performs efficient genome searching with Biostrings and BSgenome data packages. The R script loads \texttt{BSgenome.Celegans.UCSC.ce2}, which is the ce2 genome for chromosome I of Caenorhabditis elegans \cite{workflow1a}. The script finds an arbitrary nucleotide pattern in a chromosome and in an entire genome. For executing the script using RBioCloud, the job is organised into one directory, for example \texttt{BSGenome}, which contains the \texttt{GenomeSearching.R} script and all associated data. \texttt{BSGenome} also needs to contain two additional directories \texttt{Results} and \texttt{RunResults} (a similar directory structure needs to be followed for executing any job using RBioCloud). All the results that need to be generated by the script need to be directed to \texttt{Results}. \texttt{RunResults} is not submitted onto the cloud but remains on the host site to retrieve and store results of each individual run. The following sequence of five commands will execute \texttt{GenomeSearching.R} on the cloud and fetch the results onto the host site: 
\begin{itemize}
\item[1 \texttt{>}]\texttt{RBC\_GatherResource -rname `BSgenome\_instance' -rsize 1 -desc `For\_Genome\_ Searching}
\item[2 \texttt{>}]\texttt{RBC\_SubmitJob -rname `BSgenome\_instance'}
\item[3 \texttt{>}]\texttt{RBC\_ExecuteJob -rname `BSgenome\_instance' -rscript `GenomeSearching.R' -runname `Run1\_on\_BSgenome\_instance'}
\item[4 \texttt{>}]\texttt{RBC\_GetResults -rname `BSgenome\_instance' -runname `Run1\_on\_BSgenome\_instance'}
\item[5 \texttt{>}]\texttt{RBC\_TerminateResource -rname `BSgenome\_instance' -deletevol}
\end{itemize}

When the first command of the sequence is executed one EC2 instance is initialised using the Bioconductor AMI, and tagged as \texttt{BSgenome\_instance}. If optional arguments such as type of instance and EBS volume are not provided then the default values which are defined in the RBioCloud configuration file are chosen; the default values can be edited. The \texttt{BSGenome} folder is synchronised with \texttt{BSgenome\_instance} when the second command is executed; BSGenome is the current working directory from which the \texttt{RBC\_SubmitJob} is executed. The script, \texttt{GenomeSearching.R} from \texttt{BSGenome} directory is executed on \texttt{BSgenome\_instance} with a run name, \texttt{Run1\_on\_BSgenome\_instance}, when the third command is executed. The results from \texttt{Run1\_on\_BSgenome\_instance} are retrieved on to the host \texttt{Results} directory when the fourth command is executed. The Amazon resource \texttt{BSgenome\_instance} is terminated using the fifth command. The multiple execution of the \texttt{RBC\_GatherResource} command facilitates the creation of multiple instances, and multiple instances cannot have the same name. 

The job is to find nucleotide patterns in an entire genome using two methods and produce their result in two seperate files. The input is a dictionary, containing 50 patterns, each of which is a short nucleotide sequence of 15 to 25 bases. In the first method, the forward and reverse strands of seven Caenorhabditis elegans chromosomes named as chrI, chrII, chrIII, chrIV, chrV, chrX, chrM are the target. The result obtained is in a tabulated form in \texttt{ce2dict0\_ana1.txt} providing the name of the chromosome where the hit occurs, two integers giving the starting and ending positions of the hit, an indication of the hit either in the forward or reverse strand, and unique identification for every pattern in the dictionary. A sample of the output in \texttt{ce2dict0\_ana1.txt} is shown in Figure \ref{figure3} (left).

In the second method, a function which is approximately one hundred times faster is employed. One limitation of the function is that it works only when all DNA patterns searched for have a constant number of nucleotide bases. Therefore, the nucleotide patterns in the dictionary are truncated to a constant length of 15. The output of this method is also tabulated in the second result file \texttt{ce2dict0cw15\_ana2.txt} in a similar way to the first method. A sample of the output is shown in Figure \ref{figure3} (right).

\begin{figure*}
	\begin{multicols}{2}
	\footnotesize
	\begin{verbatim}
		seqname start    end      strand  patternID
		chrI    5942496	 5942511  -       pattern17
		chrI    6298363  6298377  +       pattern19
		chrI    12760564 12760587 -       pattern21
		chrI    3953136  3953150  +       pattern23
		chrI    11568996 11569018 -       pattern27
		chrI    753618   753641   +       pattern37
		...

		seqname start   end     strand  patternID
		...
		chrI    13745040        13745054        +  pattern04
		chrI    14075187        14075201        +  pattern04
		chrI    11745177        11745191        +  pattern08
		chrI    8981081 8981095 +       pattern11      
		chrI    12188778        12188792        +  pattern16
		chrI    12233665        12233679        +  pattern16
		...
	\end{verbatim}
	\end{multicols}
	\caption{Sample results obtained from first method (left) and second method (right) in \texttt{GenomeSearching.R}}
	\label{figure3}
\end{figure*}

\subsection{Test case 2: Detection of differentially expressed genes on an Instance}

The second test case is based on the \texttt{logitT} software package \cite{workflow2} available from Bioconductor\footnotemark[9]. The script executed is \texttt{logitT.R} which is a statistical method based on the Logit-t algorithm for identifying differentially expressed genes using probe-level data. The input to the script is the \texttt{spikein95} data set of the \texttt{SpikeInSubset} library \cite{spikeinsubset-1}. This data set is a subset of the Human Genome U95 data set containing a series of genes spiked-in at known concentrations and arrayed in a Latin Square format\footnote{\small \url{http://www.affymetrix.com/support/technical/sample\_data/datasets.affx}}. The Logit-t algorithm requires limited pre-processing before the actual statistical analysis and produces better results \cite{workflow2a} compared to competing approaches such the regression modelling approach \cite{statmethod-1}, the mixture model approach \cite{statmethod-2} and the Significance Analysis of Microarrays (SAM) \cite{SAM-1}. 

For executing the script using RBioCloud, the job is organised into one directory, for example \texttt{logitT}, which contains the \texttt{logitT.R} script, all associated data and the \texttt{Results} and \texttt{RunResults} directories. The following sequence of five commands will execute \texttt{logitT.R} on the cloud and fetch the results onto the host site: 

\begin{itemize}
\item[1 \texttt{>}]\texttt{RBC\_GatherResource -rname `logitT\_instance' -rsize 1 -desc `For\_Detecting\_Differentially\_Express-\newline ed\_Genes}
\item[2 \texttt{>}]\texttt{RBC\_SubmitJob -rname `logitT\_instance'}
\item[3 \texttt{>}]\texttt{RBC\_ExecuteJob -rname `logitT\_instance' -rscript `logitT.R' -runname `Run1\_on\_logitT\_instance'}
\item[4 \texttt{>}]\texttt{RBC\_GetResults -rname `logitT\_instance' -runname `Run1\_on\_logitT\_instance'}
\item[5 \texttt{>}]\texttt{RBC\_TerminateResource -rname `logitT\_instance' -deletevol}
\end{itemize}

When the \texttt{logitT.R} script is executed on the \texttt{logitT\_instance}, firstly, probe level intensities are normalised using the logit-log transformation. Then the normalised probe level intensities are standardised using Z-transformation. Student's t-tests are then performed for every Perfect Match (PM) probe in a probe set. The median t-statistic for the probe set defines Logit-t. The p-values of all the probe sets are calculated and probe sets with p-values less than 0.01 marks the detection of differentially expressed genes. The output of the algorithm is as follows:
\begin{verbatim}
	"1024_at"  "1708_at"  "32660_at" "36202_at" 
	"36311_at" "38734_at"
\end{verbatim}

\subsection{Test case 3: Normalisation of microRNA (miRNA) microarray data on a Cluster}
\begin{figure*}
	\centering
	\subfloat[Scatter plot showing on logarithmic scale array effect versus standard deviation obtained from \texttt{LVSmiRNA.R}]{\label{figure5-1}\centering \includegraphics[width=0.48\textwidth]{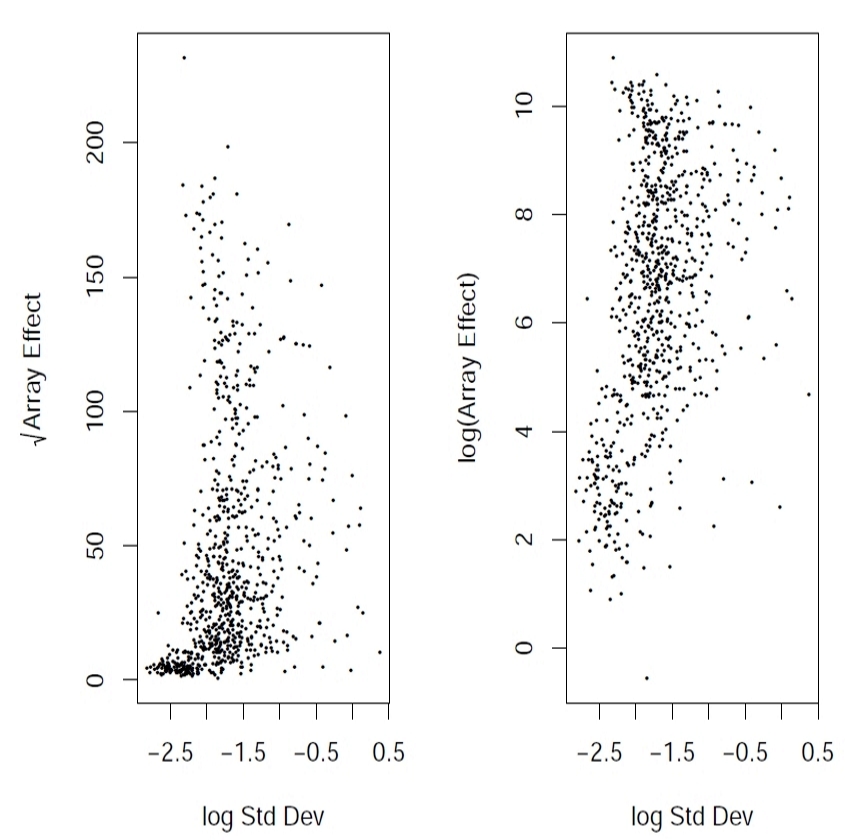}} \hspace{25pt}
	\subfloat[Box plot of miRNA data after LVS normalisation obtained from \texttt{LVSmiRNA.R}]{\label{figure5-2}\centering \includegraphics[width=0.36\textwidth]{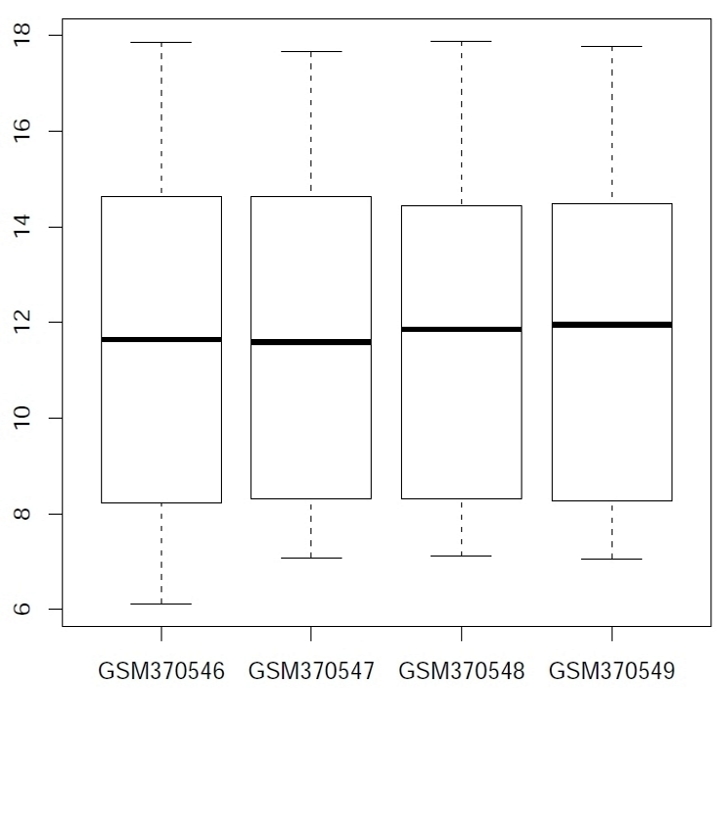}}
\caption{Results from Test case 3}
\label{figure5}
\end{figure*}

The third workflow is based on the \texttt{LVSmiRNA} software package \cite{workflow3} available from Bioconductor\footnotemark[9]. The script executed is \texttt{LVSmiRNA.R} which normalises microRNA (miRNA) microarray data. The Least-Variant Set (LVS) normalisation method \cite{workflow3a} is employed in the package and the input is the miRNA expression data \cite{workflow3b} provided as \texttt{Comparison\_Ar-\newline ray.txt}. The script then identifies a subset of miRNAs with the smallest array-to-array variation, using the \texttt{estVC} function. The first result obtained from the script is an RA-plot, which is a scatter plot (refer Figure \ref{figure5-1}) with logarithmic scales showing the array effect versus standard deviation. The second result obtained from the script is a box plot (refer Figure \ref{figure5-2}) of data after normalisation.

The \texttt{estVC} function can benefit from using parallel computation for achieving higher speed up over sequential computation, and can take a cluster object as an argument. Here Amazon clusters can come to play, and will need to be manually configured using the Amazon dashboard as shown in \cite{BioconductorCloudAMI} and \cite{workflow2c}. Employing RBioCloud will be easier as the user can configure this as a single parameter in the \texttt{RBC\_GatherResource} command. 

To execute the \texttt{LVSmiRNA.R} script on an Amazon cluster, the script and the input data needs to be provided in a directory, for example \texttt{LVSmiRNA}, and the directory also needs to contain two additional sub-directories \texttt{Results} and \texttt{RunResults}. The two graphs generated by the script needs to be directed to \texttt{Results}. \texttt{RunResults} is not submitted onto the cloud but remains on the host site to store results of every individual run. The following sequence of five commands will execute \texttt{LVSmiRNA.R} on a cloud cluster and fetch the results onto the host site: 

\begin{itemize}
\item[1 \texttt{>}]\texttt{RBC\_GatherResource -rname `LVSmiRNA\_cluster' -rsize 8 -desc `For\_LVS\_miRNA}
\item[2 \texttt{>}]\texttt{RBC\_SubmitJob -rname `LVSmiRNA\_cluster'}
\item[3 \texttt{>}]\texttt{RBC\_ExecuteJob -rname `LVSmiRNA\_cluster' -rscript `LVSmiRNA.R' -runname `Run2\_on\_LVSmiRNA\_cluster'}
\item[4 \texttt{>}]\texttt{RBC\_GetResults -rname `LVSmiRNA\_cluster' -runname `Run2\_on\_LVSmiRNA\_cluster' -frommaster}
\item[5 \texttt{>}]\texttt{RBC\_TerminateResource -rname `LVSmiRNA\_cluster' -deletevol}
\end{itemize}

A cluster with eight EC2 instances is initialised using the Bioconductor AMI, and tagged as \texttt{LVSmiRNA\_cluster} when the first command is executed. Should the optional arguments such as type of instance and EBS volume be not provided then the default values which are defined in a configuration file are chosen. The \texttt{LVSmiRNA} folder is synchronised on \texttt{LVSmiRNA\_cluster} when the second command is executed; \texttt{LVSmiRNA} is the current working directory. The script, \texttt{LVSmiRNA.R} from \texttt{LVSmiRNA} is executed on \texttt{LVSmiRNA\_cluster} with a run name, \texttt{Run2\_on\_LVSmiRNA\_cluster} when the third command is executed. The resultant graphs from \texttt{Run2\_on\_LVSmiRNA\_cluster} run is retrieved on to the host \texttt{Results} directory when the fourth command is executed. The Amazon resource \texttt{LVSmiRNA\_cluster} is terminated using the fifth command. 

\subsection{Summary}
 
\begin{figure*}
	\centering
		\includegraphics[width=0.93\textwidth]{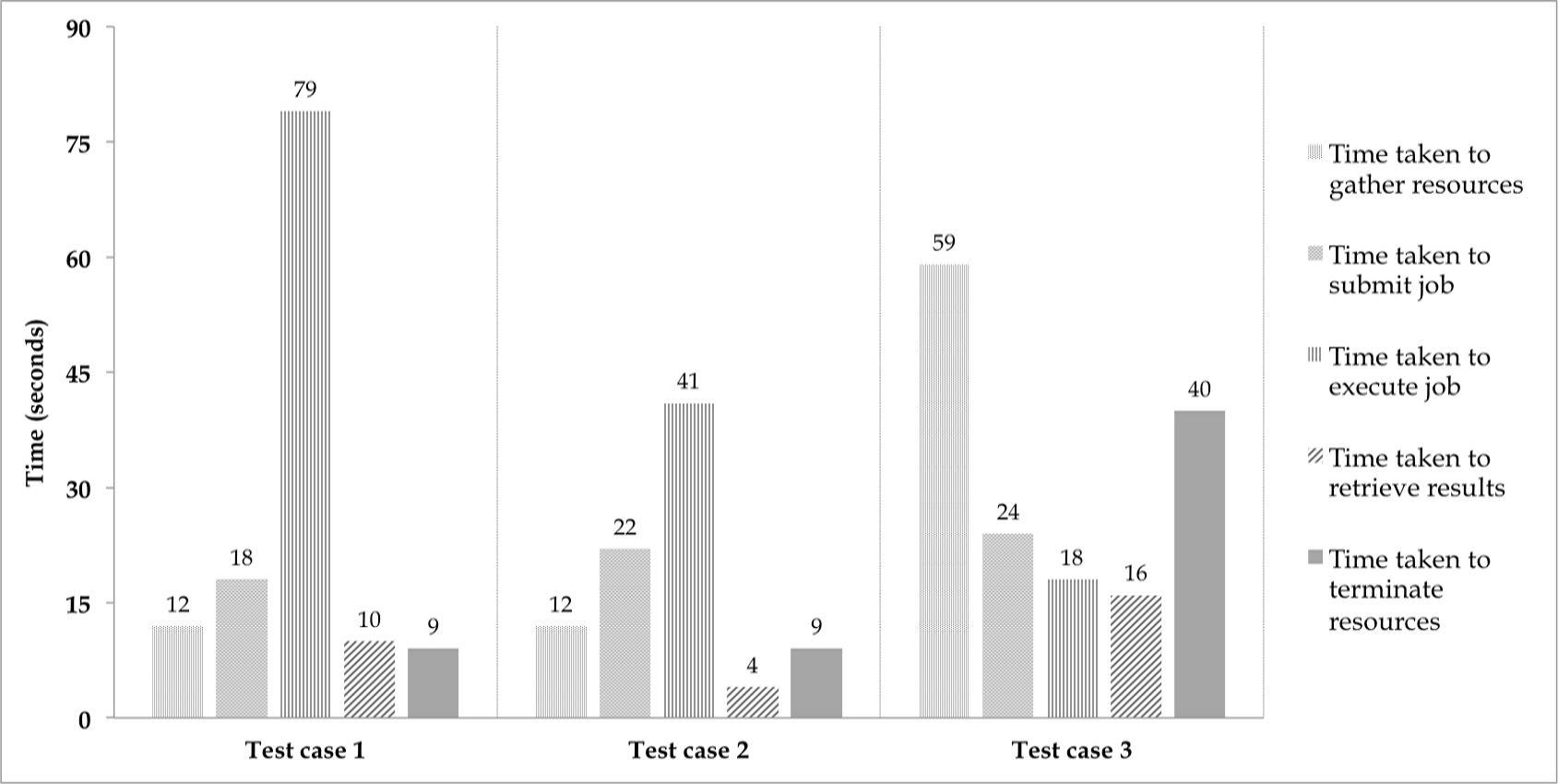}
	\caption{Time taken by the test cases on RBioCloud for gathering resources, submitting and executing jobs, retrieving results and terminating resources}
\label{figure6}
\end{figure*}

Figure \ref{figure6} shows a graph for the time taken to move data related to the job in and out of the cloud. The Amazon resources used for test case 1 and test case 2 are one m1.xlarge instance and for test case 3 is a cluster of six m1.xlarge instances. There is an increase in the time taken for initialising and terminating the cluster over the time taken for initialising and terminating one instance. Therefore, alternative techniques will need to be considered for initialising and terminating resources in parallel. This can contribute to the reduction of the overall time taken by RBioCloud. 

The time taken to submit the job is proportional to the size of the script and the input data being submitted. Large data sets required by the three test cases are available on the Amazon instances employed in this research (a custom built Amazon Machine Image (AMI) based on the Bioconductor AMI is used in this research). Genome searching takes 79 seconds and the detection of differential expression of genes takes 41 seconds to complete execution. The potential for parallelism in these jobs and scaling the job across multiple instances need to be explored to achieve speed up. The third test case exploits parallelism and executes on a cluster of six instances taking 18 seconds for completing the job. Again the time for retrieving results is proportional to the size of the files produced as results. The second test case takes the least time for retrieval since it produces a small output. 

Additional test cases to confirm the feasibility of RBioCloud were performed using a large number of scripts provided by over 150 Bioconductor packages. For example, the following were performed using RBioCloud on a large-memory Amazon EC2 instance: 
\begin{itemize}
\item[(a)] Estimation of False Discovery Rate (FDR) using Significance Analysis of Microarrays (SAM) \cite{SAM-1} and the Empirical Bayes Analyses of Microarrays (EBAM) \cite{EBAM-1} provided as \texttt{siggenes.R} available from the \texttt{siggenes} package \cite{siggenes-1}, 
\item[(b)] Joint Deregulation Analysis (JODA) for quantifying changes due to regulation of genes between two distinct cell populations provided as \texttt{JodaVignette.R} available from the \texttt{joda} package \cite{JODA-1}, and 
\item[(c)] Analysing ChIP-seq data including the detection of protein-bound genomic regions provided as \texttt{CSAR.R} available from the \texttt{CSAR} package \cite{CSAR-1}. 
\end{itemize}

One observation from the test cases is that the full advantage of the cloud is exploited when jobs harness the potential of parallelism. The results obtained from the additional test cases are beyond the scope of this paper and will be reported elsewhere.

\section{Conclusions}
\label{conclusions}
Gathering and managing vast cloud resources in the computational biology or bioinformatics setting for executing an analytical job can be cumbersome. This is not because cloud resources aren't readily accessible, but the pipeline for executing an analytical job on the cloud requires extensive knowledge of the cloud. While high-performance computer architects may be able to design and deploy such workflows for production based applications it may not be easily possible for biologists with limited high-performance computing skills to perform ad hoc analytics. To allow analytical jobs to fully benefit from the cloud there needs to be a framework that can seamlessly adapt analytical jobs located on a host site for execution on the cloud, provide minimal difference between a personal desktop and the cloud, and offer data and resource management easily on the cloud.

In this paper, such a framework, `RBioCloud', which is light-weight and easily deployable has been designed and developed to support analytical jobs comprising R scripts which employ Bioconductor packages. The framework is deployed between a host site and the cloud, and a set of five command line tools are offered for analytical workflows to facilitate gathering resources, submitting a job, executing a job, retrieving results, and terminating resources. The research contributions of RBioCloud has been a framework to (i) seamlessly handle a diverse range of analytical job on the cloud, (ii) abstract the complexities of cloud set up and configuration, (iii) easily access and manage cloud resources, thereby saving time of domain scientists, and (iv) remotely access cloud resources from a workstation with seemingly minimal differences. Test cases using Bioconductor and R-based jobs demonstrate the feasibility of RBioCloud. Three test cases have been employed to validate the feasibility of RBioCloud. In the first test case, genome searching, and in the second test case, detection of differential expression of genes were both performed on a single Amazon EC2 instance. In the second test case, normalisation of microRNA (miRNA) microarray data was performed using a cluster of Amazon EC2 instances. The framework is available for download from \texttt{\url{http://www.rbiocloud.com}}.

Future efforts will be made towards extending RBioCloud for dynamic and automated management of compute and storage resources on the cloud and submission and execution of multiple jobs. On top of on-demand instances which are available for fixed price the cost effective solution of bidding for spare instances will be explored.

\end{document}